\documentclass[
  aps,
  pra,
  reprint,            
  superscriptaddress, 
  amsmath,amssymb,
  notitlepage
]{revtex4-2}

\usepackage[utf8]{inputenc}
\usepackage[T1]{fontenc}

\usepackage{hyperref}       
\usepackage{url}            
\usepackage{booktabs}       
\usepackage{amsfonts}       
\usepackage{nicefrac}       
\usepackage{microtype}      

\usepackage{graphicx}       
\graphicspath{{media/}}     

\usepackage[normalem]{ulem}
\usepackage{array}
\usepackage{tikz}
\usetikzlibrary{positioning,quotes}
\usepackage{subcaption}     
\usepackage{xcolor}
\usepackage{soul}
\usepackage{amsthm}
\usepackage{framed}
\usepackage{mathtools}
\usepackage{quantikz}
\usepackage{cancel}

\usepackage{algorithm}
\usepackage{algpseudocode}

\theoremstyle{plain}

\theoremstyle{definition}

\theoremstyle{definition}

\usepackage{xcolor}

\usepackage[utf8]{inputenc}
\usepackage{hyperref}
\usepackage{setspace} 
\usepackage{graphicx}

\begin{document}







\author{Bora Baran}
\email{b.baran@fz-juelich.de}
\affiliation{Forschungszentrum Jülich, Institute of Quantum Control,%
 Peter Grünberg Institut (PGI-8), 52425 Jülich, Germany}
\affiliation{Institute for Theoretical Physics, University of Cologne, Zuelpicher Strasse 77, 50937 Cologne, Germany}

\author{Tommaso Calarco}
\affiliation{Forschungszentrum Jülich, Institute of Quantum Control,%
 Peter Grünberg Institut (PGI-8), 52425 Jülich, Germany}
\affiliation{Institute for Theoretical Physics, University of Cologne, Zuelpicher Strasse 77, 50937 Cologne, Germany}
\affiliation{Dipartimento di Fisica e Astronomia, Università di Bologna, 40127 Bologna, Italy}

\author{Matthias M. Müller}
\affiliation{Forschungszentrum Jülich, Institute of Quantum Control,%
 Peter Grünberg Institut (PGI-8), 52425 Jülich, Germany}

\author{Felix Motzoi}
\affiliation{Forschungszentrum Jülich, Institute of Quantum Control,%
 Peter Grünberg Institut (PGI-8), 52425 Jülich, Germany}
\affiliation{Institute for Theoretical Physics, University of Cologne, Zuelpicher Strasse 77, 50937 Cologne, Germany}




\title{Interaction-resolved decomposition of multi-qubit unitaries\\ via computational-basis phases}

\begin{abstract}
\noindent
In multi-qubit quantum control, target unitary operations are commonly specified through full-unitary target descriptions and assessed through global comparison measures. In this work, we introduce an interaction-resolved decomposition of n-qubit unitaries that provides explicit access to their many-body interaction structure through computational-basis phases collected in a diagonalizing frame. Such a frame is conveniently given by local rotations for many operationally relevant operations, including gates generated by single Pauli strings or commuting sets of Pauli strings, such as stabilizer operations, controlled-phase gates, Toffoli-type operations, and Ising interactions. We derive parity-weighted sums of these computational-basis phases that exactly and uniquely resolve k-body interaction terms supported on arbitrary qubit subsets, which we term support-selective phase invariants. These invariants provide an interaction-resolved coordinate system that organizes unitary operations according to their multipartite interaction structure, giving direct access to local, pairwise, tripartite, and general $k$-partite interaction content underlying entangling operations. This enables the formulation of selective quantum optimal control targets for synthesizing desired combinations of many-body interactions. We supplement this with numerical demonstrations for a representative hardware model, a realistic nitrogen-vacancy spin register, where we synthesized isolated tripartite interactions up to local equivalence within a single control pulse, guided by these invariants, for both diagonal (ZZZ) and non-diagonal (XZZ) terms.
\end{abstract}

\maketitle

\section*{Introduction}

Quantum optimal control provides systematic methods for shaping external fields to steer a quantum system through a desired transformation, enabling precise implementation of logical gates, state transfers, and error-resilient operations \cite{koch2022quantum}. As a result, quantum optimal control has become a central tool for engineering high-performance operations across a wide range of quantum hardware platforms, including trapped ions \cite{monz201114}, neutral-atom arrays \cite{evered2023high, levine2019parallel, omran2019generation}, superconducting circuits \cite{li2024hardware, zhao2025microwave, barnes2017fast, li2024experimental, gao2025ultrafast}, and solid-state spin qubits such as nitrogen-vacancy centers and related platforms \cite{dong2024time, ChouNVOptim, wang2025circuit}. These experimental developments are supported by a mature theoretical and algorithmic framework, including gradient-based and variational optimization methods that enable systematic improvement of control fields \cite{Grace10112007, sauvage2022optimal, khaneja2005optimal, motzoi2011optimal, caneva2011chopped, rach2015dressing, doria2011optimal, muellercontrol}.
Established approaches such as GRAPE \cite{khaneja2005optimal, de2011second, motzoi2011optimal, hogben2011spinach}, Krotov \cite{konnov1999deterministic, Reich07062014,goerz2019krotov,reich2012monotonically,goerz2014optimal}, and dCRAB \cite{caneva2011chopped, rach2015dressing, doria2011optimal, muellercontrol} formulate quantum optimal control as the minimization of a cost functional that quantifies the deviation between the realized and desired evolution, independent of the specific choice of controls \cite{glaser2015training,khaneja2005optimal, preti2022continuous, barnes2017fast, palao2013steering, reich2012monotonically, caneva2011chopped,  brif2010control, zhang2003geometric, goerz2014optimal, theis2018counteracting, guery2019shortcuts,  calzavara2025classical}. In these frameworks the objective is typically defined by specifying a target unitary transformation.

\medskip
\noindent
However, conventional full-unitary target descriptions and associated global comparison measures do not explicitly resolve the many-body interaction structure underlying a realized operation or its deviation from a target \cite{khaneja2005optimal,motzoi2011optimal, zhang2003geometric, goerz2015optimizing}. In multi-qubit systems, where operations generally comprise combinations of interaction terms supported across multiple qubit subsets, this limits the direct assessment and targeting of specific many-body terms. These considerations motivate the development of interaction-resolved representations that expose individual interaction terms and enable control objectives formulated directly in terms of the desired many-body structure of a quantum operation. Such capabilities are particularly relevant for stabilizer-based quantum error correction and related settings, where specific multi-qubit parity operators and interaction structures play a central operational role but are typically realized through decompositions into lower-order gates and ancilla-mediated constructions rather than direct synthesis~\cite{gottesman1997stabilizer,fowler2012surface,litinski2019game,taminiau2014universal,cramer2016repeated}.

\medskip
\noindent
In this work we introduce an interaction-resolved decomposition of multi-qubit unitary transformations that provides explicit access to their many-body interaction structure. The key idea is to work in a diagonalizing frame, where the unitary is fully characterized by the phases acquired by computational-basis states, which can be accessed without requiring full tomography through phase-sensitive measurements \cite{childress2006coherent, balasubramanian2009ultralong}. This framework is especially practical for a broad class of operationally relevant multi-qubit operations, including gates generated by single Pauli strings or commuting sets of Pauli strings, such as stabilizer operations, controlled-phase gates, Toffoli-type gates, because for those operations diagonalizing frames can be obtained through local transformations. From the computational-basis phases, we construct parity-weighted sums that exactly and uniquely resolve $k$-body interaction terms supported on arbitrary qubit subsets, while remaining invariant under terms supported on complementary subsets; we refer to these constructions as \emph{support-selective phase invariants}. These invariants provide an interaction-resolved coordinate system that enables the direct assessment of many-body interaction terms and the formulation of interaction-specific control objectives for synthesizing prescribed combinations of interaction terms. This, furthermore, induces generalized equivalence classes obtained by requiring arbitrary selected $k$-body interaction terms to coincide, with local equivalence recovered as a special case.


\medskip
\noindent
Within this phase-invariant-based decomposition, one-body terms correspond to local phases, while support-selective phase invariants associated with subsets $S$ with $|S|\geq 2$ capture nonlocal many-body interaction terms. In this sense, the construction extends the idea of local invariants used in Cartan/KAK-based characterizations of two-qubit gates~\cite{watts2015optimizing,goerz2015optimizing,zhang2003geometric,rezakhani2004characterization,frey2026optimal,muller2011optimizing} by resolving interaction terms supported on arbitrary qubit subsets and across arbitrary $k$-body subclasses of gates. Rather than characterizing a gate only up to local equivalence, the resulting decomposition provides direct access to its general underlying many-body interaction structure, including local, pairwise, tripartite, and general \(k\)-partite interaction content underlying multipartite entangling operations.

\medskip
\noindent
To illustrate the utility of this framework for realistic quantum hardware, we consider the nitrogen-vacancy (NV$^-$) center in diamond as a representative solid-state spin platform. Its electronic spin, coupled to nearby nuclear spins, exhibits stable hyperfine-mediated interactions that naturally support multi-spin conditional control \cite{wang2025circuit, phila, Chen, Schir, neumann, rembold2020introduction}. Using interaction-specific control objectives formulated through support-selective phase invariants, we demonstrate in realistic simulations of an NV spin register the synthesis of isolated tripartite interaction terms up to local equivalence, including both diagonal and non-diagonal cases, within a single control pulse.

\section{Interaction-resolved Decomposition}
\label{sec:interaction_resolved_representation}

\subsection{Computational-Basis Representation of Multi-Qubit Unitaries}
\label{sec:comp_frame}
\noindent
To resolve the interaction structure of multi-qubit operations we begin by considering the representation of unitary transformations in a diagonalizing basis. Many multi-qubit gates of practical relevance are generated by Pauli-strings $P_S \in \{I,X,Y,Z\}^{\otimes n}$, so that a generator can be written as
\begin{equation}
G = \sum_{S\subseteq\{1,\dots,n\}} \theta_S P_S,
\label{eq:generator_theta}
\end{equation}
where each $\theta_S$ are the $k$-body interaction strengths of a Pauli-string $P_S$ and $S$ corresponds to a set of qubit indices on which a Pauli-string $P_S$ acts,  with $k=|S|$. The generator $G$ produces unitaries of the form
\begin{equation}
U = e^{-iG}.
\end{equation}
For every Pauli string $P_S$ there exists a unitary $U_P$ such that
\begin{equation}
U_P P_S U_P^\dagger =  Z_S,
\end{equation}
where 
\begin{equation}
Z_S = \prod_{i\in S} Z_i
\end{equation}
denotes a Pauli-$Z$ string supported on a subset $S \subseteq \{1,\dots,n\}$.\\

\noindent
We are particularly interested in unitary gates generated by single Pauli-string or mutually commuting sets of Pauli-strings: in such cases, the same local transformation simultaneously diagonalizes all generator components, yielding 
\begin{equation}
G_{\mathrm{diag}}= U_P G U_P^\dagger = \sum_{S\subseteq\{1,\dots,n\}}\, \theta_S Z_S.
\label{eq:diag_generator}
\end{equation}
The corresponding unitary takes the form
\begin{equation}
    U_{\mathrm{diag}} = \exp\!\left(
    -i \sum_{S\subseteq\{1,\dots,n\}}\, \theta_S Z_S \right).
    \label{eq:diag_unitary}
\end{equation}

If $U_P$ is a local transformation (e.g., for a single Pauli string), then the $Z_S$ correspond exactly to $k$-body interactions on the subset $S$ and then reveal the many-body interaction structure encoded in $U$.

\subsection{Resolving Interaction Contributions via Computational-Basis Phases}

\noindent
Given a unitary $U_{\mathrm{diag}}$ expressed in a diagonalizing frame, such as in Eq.~\eqref{eq:diag_unitary}, we show that any Pauli-$Z$ interaction strength $\theta_S$ can be can recovered by linear combinations of phases acquired by computational-basis states under the application of $U_{\mathrm{diag}}$. \\

\noindent
Since $U_{\mathrm{diag}}$ is diagonal in the computational basis, its action on basis states
$\ket{\vec{x}}$, with $\vec{x}\in\{0,1\}^n$, is
\begin{equation}
U_{\mathrm{diag}}\ket{\vec{x}}
=
e^{-i\phi(\vec{x})}\ket{\vec{x}},
\end{equation}
where $\phi(\vec{x})$ denotes the phase acquired by the basis state $\ket{\vec{x}}$. The full unitary can therefore be written as
\begin{equation}
U_{\mathrm{diag}}
=
\sum_{\vec{x}\in\{0,1\}^n}
e^{-i\phi(\vec{x})}
\ket{\vec{x}}\bra{\vec{x}},
\label{eq:diag_unitary_comp_phases}
\end{equation}
so that we write its generator in terms of computational-basis phases, 
\begin{equation}
\Phi
=
\sum_{\vec{x}\in\{0,1\}^n} \phi(\vec{x})\,|\vec{x}\rangle\langle\vec{x}|,
\quad \text{with} \quad
U_{\mathrm{diag}} = e^{i\Phi}.
\label{eq:phi_generator}
\end{equation}
Since $\Phi$ is diagonal in the computational basis, it admits a natural expansion in the diagonal Pauli-$Z$ operator basis
\(Z_S = \prod_{i\in S} Z_i,\,\text{with}\, S \subseteq \{1,\dots,n\}, \)
so that it can also be written as a unique combination of $Z_S$ terms
\begin{equation}
\Phi
=
\sum_{S\subseteq\{1,\dots,n\}}
\phi(S)\, Z_S ,
\label{eq:phase_pauli_expansion}
\end{equation}
where $\phi(S)$ are linear combinations of the coefficients $\phi(\vec{x})$ in Eq.\eqref{eq:phi_generator}. \\
Since Eqs.~\eqref{eq:diag_generator} and \eqref{eq:phase_pauli_expansion} provide two equivalent representations of the same diagonal generator in the Pauli-$Z$ basis, the corresponding expansion coefficients must coincide,
\begin{equation}
\phi(S)=\theta_S.
\label{eq:phi_theta_equivalence}
\end{equation}
The explicit form of the coefficients follows from projection onto the Pauli-$Z$ basis using the trace inner product,
\begin{equation}
\phi(S)
=
2^{-n}\mathrm{Tr}(\Phi Z_S)
=
2^{-n}
\sum_{\vec{x}\in\{0,1\}^n}
\langle\vec{x}|\Phi Z_S|\vec{x}\rangle.
\label{eq:phase_invariant}
\end{equation}
Since computational-basis states are eigenstates of Pauli-$Z$ strings,
\begin{equation}
Z_S|\vec{x}\rangle
=
(-1)^{\sum_{i\in S}x_i}|\vec{x}\rangle,
\end{equation}
the coefficients reduce to
\begin{equation}
\phi(S)
=
2^{-n}
\sum_{\vec{x}\in\{0,1\}^n}
(-1)^{\sum_{i\in S}x_i}
\langle\vec{x}|\Phi|\vec{x}\rangle.
\end{equation}
Using
\(
\langle\vec{x}|\Phi|\vec{x}\rangle=\phi(\vec{x}),
\)
we obtain
\begin{equation}
\phi(S)
=
2^{-n}
\sum_{\vec{x}\in\{0,1\}^n}
(-1)^{\sum_{i\in S}x_i}
\phi(\vec{x}).
\label{eq:phase_coefficients_parity_sum}
\end{equation}

\medskip
\noindent
Thus the coefficients $\theta_S$ can be uniquely obtained as parity-weighted sums of computational-basis phases. As this returns a specific many-body interaction contribution strength $\theta_S$ by combining phases, while canceling other contributions due to basis-orthogonality, we refer to the construction $\phi(S)$ in Eq.~\eqref{eq:phase_coefficients_parity_sum} as \emph{support-selective phase invariants}.

\medskip
\noindent
\textbf{Remark.}
The coefficients $\phi(S)$ coincide mathematically with the Walsh-Hadamard transform of the computational-basis phases on the Boolean hypercube $\{0,1\}^n$~\cite{welch2014efficient}; here they emerge from projecting the unitary generator onto the Pauli-$Z$ operator basis.\\

\noindent
The phase invariants $\phi(S)$~\eqref{eq:phase_coefficients_parity_sum} provide an interaction-resolved coordinate system for unitaries in a diagonalizing frame, as they decompose any diagonalized unitary into $k$-body contributions. The diagonal unitary can therefore be written as
\begin{equation}
U_{\mathrm{diag}}
=
\exp\!\left(-
i\sum_{S\subseteq\{1,\dots,n\}}
\phi(S)\, Z_S
\right),
\label{eq:phi_expansion_u_diag}
\end{equation}
where the coefficients $\phi(S)$ specify the many-body contribution strength supported on the subset $S$.

\subsection{Equivalence Classes}
\label{sec:interaction_equivalence_classes}

\noindent
Since the support-selective phase invariants $\phi(S)$ provide an interaction-resolved coordinate system for unitaries through Eq.~\eqref{eq:phi_expansion_u_diag}, they naturally induce equivalence relations by specifying which interaction coordinates are required to coincide.

For a fixed support subset $S$, two unitaries $U$ and $V$, represented in diagonalized form, are said to be $S$-interaction equivalent if they possess the same coordinate along the interaction direction $Z_S$:
\begin{equation}
\phi_U(S)
=
\phi_V(S)
\quad \mathrm{mod}\; 2\pi.
\end{equation}

More generally, for a chosen collection of supports
\[
\mathcal{A}\subseteq 2^{\{1,\dots,n\}},
\]
we define the interaction-equivalence relation
\begin{equation}
U \sim_{\mathcal{A}} V
\quad\Longleftrightarrow\quad
\phi_U(S)=\phi_V(S)
\; \mathrm{mod}\; 2\pi,
\qquad
\forall S\in\mathcal{A}.
\end{equation}

Thus, each selected set of supports $\mathcal{A}$ determines a corresponding equivalence class on \(U(2^n)\) via the interaction-coordinate space: unitaries belonging to the same equivalence class agree on the specified many-body interaction coordinates while remaining free to differ along coordinates associated with subsets outside $\mathcal{A}$.

In particular, choosing
\[
\mathcal{A}_{\geq2}
=
\left\{
S\subseteq\{1,\dots,n\}
:\ |S|\geq2
\right\}
\]
fixes all nonlocal interaction coordinates while allowing arbitrary one-body coordinates. This recovers an analogue of local equivalence within the interaction-resolved decomposition: local phases may vary, but the nonlocal interaction-coordinate structure is preserved.\\

\noindent
Conversely, selecting a single support $S$ defines a more flexible equivalence class containing all unitaries sharing the same $S$-body interaction coordinate, independent of all remaining coordinates.\\

\noindent
These equivalence classes are useful for quantum control because they allow objectives to be formulated at different levels of interaction resolution within the same coordinate framework. Rather than matching a full target unitary, one may require only membership in a prescribed interaction-equivalence class, for example by constraining selected interaction coordinates $\phi(S)$ while leaving local or otherwise irrelevant coordinates unconstrained.


\section{Interaction-resolved Quantum Optimal Control}
\label{sec:results}

\noindent
\noindent
To demonstrate the practical utility of the interaction-resolved decomposition we apply it to formulate interaction-resolved control objectives based on support selective phase invariants, we apply the approach on a representative and realistic platform, a solid-state spin register based on a nitrogen-vacancy (NV) center in diamond. For a given control pulse, the system is evolved under the effective Hamiltonian to obtain the propagator, from which computational-basis phases are extracted in the corresponding diagonalizing frame and combined into interaction-resolving support-selective phase invariants used as optimization targets.

\medskip
\noindent
We present two illustrative examples. In both, we consider a three-qubit entangler, demonstrating how how phase-invariant control objectives selectively guide the desired interaction structure of a unitary operation to synthesize a prescribed many-body interaction structure. We then synthesize a non-diagonal three-qubit entangler by performing the optimization in a suitable diagonalizing frame, demonstrating that the same framework extends naturally to unitaries generated by non-diagonal Pauli strings.

\subsection{Control Objectives in Interaction-Resolved Coordinates}
\label{sec:control_objective_phase_invariants}

\noindent
Target unitary operations can be specified by assigning desired values $\{\phi_S^\star\}$ to support-selective phase invariants associated with desired many-body interaction terms, so that the desired structure is prescribed for the unitary gate as given in Eq.~\eqref{eq:phi_expansion_u_diag}.


\noindent
To quantify deviations from desired many-body interaction contributions, a distance measure can be defined on the support-selective phase invariants, circumventing the need to use a global distance measure.\\
Given target values for a chosen set of qubit subsets  $\mathcal{S}_{\mathrm{target}}$, we define the interaction-resolved control cost function
\begin{equation}
\label{eq:phase-invariant-cost}
\mathcal{J}
=
\sum_{S\in\mathcal{S}_{\mathrm{target}}}
w_S
\Bigl[
1-\cos\!\bigl(2(\phi(S)-\phi_S^\star)\bigr)
\Bigr],
\end{equation}
where $w_S$ determines the relative importance of the corresponding interaction terms. 
Furthermore, this formulation also enables interaction-selective optimization: only the interaction contributions associated with subsets in $\mathcal{S}_{\mathrm{target}}$ enter the objective function. For example, leaving degrees of freedom, such as local terms, unconstrained, enlarges the space of admissible solutions, allowing the optimization to exploit this flexibility to find more efficient implementations.

\medskip
\noindent
\textbf{Remark.}
The factor of two in Eq.~\eqref{eq:phase-invariant-cost} reflects the $\pi$-periodicity of Pauli-$Z$ interaction phases. Indeed,
\begin{equation}
e^{-i(\theta+\pi)Z_S}
=
-\,e^{-i\theta Z_S},
\label{eq:pi_peridicity}
\end{equation}
so shifting the interaction phase by $\pi$ changes only the collective sign of the operator without modifying its physical action. The objective therefore identifies interaction phases differing by integer multiples of $\pi$ as physically equivalent.

\subsection{Model Spin Hamiltonian}
\label{sec:spin_model}

The following Hamiltonian model, describing the dynamics of a spin register based on a nitrogen-vacancy (NV) center in diamond, serves as the working model for the simulated platform, for which we optimize the controls. A full derivation of this Hamiltonian starting from the laboratory-frame NV Hamiltonian is given in Appendix~\ref{app:nv_model}. In this platform an electronic spin is coupled to nearby nuclear spins, forming a controllable multi-qubit system driven by microwave control fields. In the interaction picture with respect to the free precessing parts of the  Hamiltonian and after applying the rotating-wave approximation (RWA) at the electronic resonance, the Hamiltonian takes the form
\begin{equation}
\begin{aligned}
&\notag H_{\mathrm{NV}}(t)\\
&=\frac{\Omega(t)}{2}\sum_{\mathbf m}
\big[\sigma_x\cos((\omega_{\mathrm{MW}}-\Lambda(\mathbf m))\,t)\\
&\hspace{1.0cm}-\sigma_y\sin((\omega_{\mathrm{MW}}-\Lambda(\mathbf m))\,t)\big] \otimes \ket{\mathbf m}\!\bra{\mathbf m} \\
&-\;\frac{\Omega(t)}{2}\big[\sigma_x\sin((\omega_{\mathrm{MW}}-\Lambda_s) t)+\sigma_y\cos((\omega_{\mathrm{MW}}-\Lambda_s) t)\big] \\
&\hspace{1.0cm}\otimes\sum_i \theta_i \big[I'_{ix}\cos(\delta_i t-\phi_i)+I'_{iy}\sin(\delta_i t-\phi_i)\big],
\end{aligned}
\label{eq:H_NV_RWA_main}
\end{equation}
The first term describes the coherent microwave drive of the electronic qubit conditioned on the nuclear configuration, while the second term captures the weak microwave-mediated coupling to nearby nuclear spins.  
Here, $\Omega(t)$ is the Rabi frequency that can be tuned via a microwave magnetic field (with carrier frequency $\omega_{\mathrm{MW}}$) and serves as our control pulse, $\Lambda_s= D - \gamma_e B_0$ describes the bare electron transition frequency between its two ground-state levels \(\ket{0_e}\!\leftrightarrow\!\ket{-1_e}\), including zero-field splitting $D$ and the Zeeman shift due to the static magnetic field $B_0$ with the gyromagnetic ratio $\gamma_e$. $\Lambda(\mathbf{m})$ describes the configuration-dependent transition frequencies of the electron, with  $\mathbf m=(m_N,\{m_{C_i}\})$ labeling each nuclear spin register configuration, i.e.\ a joint eigenstate of the nuclear spin operators along the quantization axis, where the quantum numbers $m_N$ and $m_{C_i}$ correspond to the eigenvalues of the spin operator $I_z$ for the $^{14}$N and $^{13}$C nuclei. 
The quantities $\delta_i= \frac{\gamma_i B_0 + \omega_i}{2}$ describe weak effective transverse modulations of the nuclear spin. The parameters $\theta_i$ and $\phi_i$ characterize the local hyperfine misalignment of the nuclear spins.\\
The specific physical parameters used in the numerical simulations of the NV register are given in Appendix~\ref{app:nv_params}.

\subsection{Pulse Parametrization and Optimization}
\label{sec:pulse_parametrization} 
We control the system by optimizing the temporal shape of the Rabi frequency appearing in Eq.~\eqref{eq:H_NV_RWA_main}. We start by the parametrization 
\begin{equation}
\Omega(t) = \chi(t)\sum_{i=1}^{N_c} a_{i} \cos(\omega_{c,i} t + \phi_{c,i}),
\label{eq:pulse_parametrization}
\end{equation}
with frequency amplitudes $a_{i}$ carrying the physical units of the control envelope, control frequencies $\omega_{c,i}$, and control phases $\phi_{c,i}$.
The resulting parameters are collected into a time-independent parameter 
\begin{equation}
\mathbf p =
(a_{1},\ldots,a_{N_c},\;
\omega_{c,1},\ldots,\omega_{c,N_c},\;
\phi_{c,1},\ldots,\phi_{c,N_c}),
\label{eq:pulse_parametrization_example}
\end{equation}
which are to be optimized iteratively. All optimized pulse parameters, resulting in the following, are provided in the public repository~\cite{baran2026_multiqubit_control}.
The multiplicative factor $\chi(t)$ is a Tukey (tapered cosine) window that smoothly turns the pulse on and off,
\begin{equation}
\chi(t)=\tfrac12\left[1-\cos\!\left(2\pi\frac{t}{T}\right)\right], 
\qquad t\in[0,\alpha T],
\end{equation}
The parameter $\alpha\in[0,1]$ controls the taper fraction, we generally set it to $\alpha=0.15$.

\noindent
This analytic parametrization~\eqref{eq:pulse_parametrization}, used, similarly, for example in CRAB~\cite{caneva2011chopped,rach2015dressing}, GRAPE~\cite{khaneja2005optimal,de2011second}, and hybrid approaches~\cite{motzoi2011optimal,sorensen2018quantum,sorensen2020optimization,preti2022continuous}, provides a compact and experimentally compatible representation of the control fields.

Thus, for a given parameter vector $\mathbf p$, the control pulse
$\Omega(t;\mathbf p)$ defines the effective propagator
\begin{equation}
U_{\mathrm{pulse}}(\mathbf p,T)
=
\mathcal{T}
\exp\!\left(
-i\int_0^T
H_{\mathrm{NV}}(\Omega(t;\mathbf p),t)\,dt
\right).
\label{eq:U_pulse_definition}
\end{equation}

\subsection{Diagonal Three-Qubit Entangler}
\label{sec:results_nv_ccz}

\noindent
We first demonstrate the phase-invariant control framework by synthesizing a diagonal three-qubit entangling gate. The target operation is
\begin{equation}
\label{eq:desired_u_diag}
U_{\mathrm{target}} = e^{\,i\frac{\pi}{4} Z\!\otimes\!Z\!\otimes\!Z},
\end{equation}
which produces genuine tripartite entanglement and generates a GHZ state up to local phases~\cite{coffman2000distributed,dur2000three}.\\

\noindent
The system dynamics are described by the effective Hamiltonian in Eq.~\eqref{eq:H_NV_RWA_main}, and the microwave control fields are parameterized according to Eq.~\eqref{eq:pulse_parametrization_example}. We consider the logical subsystem of an electronic spin ($A$) coupled to two proximal $^{13}$C nuclear spins ($B$ and $C$), while the $^{14}$N nuclear spin is prepared in a fixed $m_I$ state and acts as a spectator. Within the logical $A\!\otimes\!B\!\otimes\!C$ subspace the optimization seeks to realize the target gate~\eqref{eq:desired_u_diag} using phase-invariant control objectives. We therefore consider the projected propagator $U^{A\otimes B\otimes C}_{(\mathrm{pulse})}(\mathbf p,T)$, corresponding to the restriction of the full propagator $U_{\mathrm{pulse}}(\mathbf p,T)$ defined in Eq.~\eqref{eq:U_pulse_definition} to the logical $A\!\otimes\!B\!\otimes\!C$ subspace.\\

\noindent
Since the target gate is already diagonal, no diagonalizing transformation is required and the projected propagator $U^{A\otimes B\otimes C}_{(\mathrm{pulse})}(\mathbf p,T)$ can directly be treated in the computational basis
\begin{equation}
U^{A\otimes B\otimes C}_{(\mathrm{pulse})}(\mathbf p,T)
=
\sum_{a,b,c\in\{0,1\}} e^{-i\phi_{abc}}\ket{abc}\bra{abc},
\label{eq:Uzzzphase}
\end{equation}
from which the many-body interaction contribution strength are extracted via the phase invariants $\phi(S)$ constructed from the phases $\phi_{abc}$. For a three-qubit system the invariants are
\begin{equation}
\begin{aligned}
&\phi(\{a\}),\ \phi(\{b\}),\ \phi(\{c\}),\quad
\phi(\{a,b\}),\ \phi(\{a,c\}),\ \phi(\{b,c\}),\\
&\phi(\{a,b,c\}).
\end{aligned}
\end{equation}
Explicit expressions of the phase invariants in terms of the computational-basis phases $\{\phi_{abc}\}_{a,b,c\in\{0,1\}}$ are given in Appendix~\ref{app:explicit_3q_invariants}. \noindent
The control objective is defined by specifying the target interaction phases
\begin{equation}
\begin{aligned}
&\phi^\star(\{a,b,c\}) = \tfrac{\pi}{4}, \\
&\phi^\star(\{a,b\})
=
\phi^\star(\{a,c\})
=
\phi^\star(\{b,c\})
= 0,
\end{aligned}
\label{eq:Uzzzinvariants}
\end{equation}
while leaving the single-qubit phase invariants unconstrained, as they correspond to locally correctable degrees of freedom. The optimization minimizes the $\pi$-periodic cost function $\mathcal{J}_{\mathrm{3q}}$ defined in Eq.~\eqref{eq:phase-invariant-cost}, with weights chosen based on relative importance and such that only nonlocal interactions ($|S|\ge2$) contribute: $w(\{{a,b,c}\})=0.5$, $w(|S|=2)=0.2$, $w(|S|=1)=0.0$. Additional regularization terms penalize leakage, non-unitarity, and rapid control variations.

\begin{figure}[h!]
\centering
\includegraphics[width=0.47\textwidth]{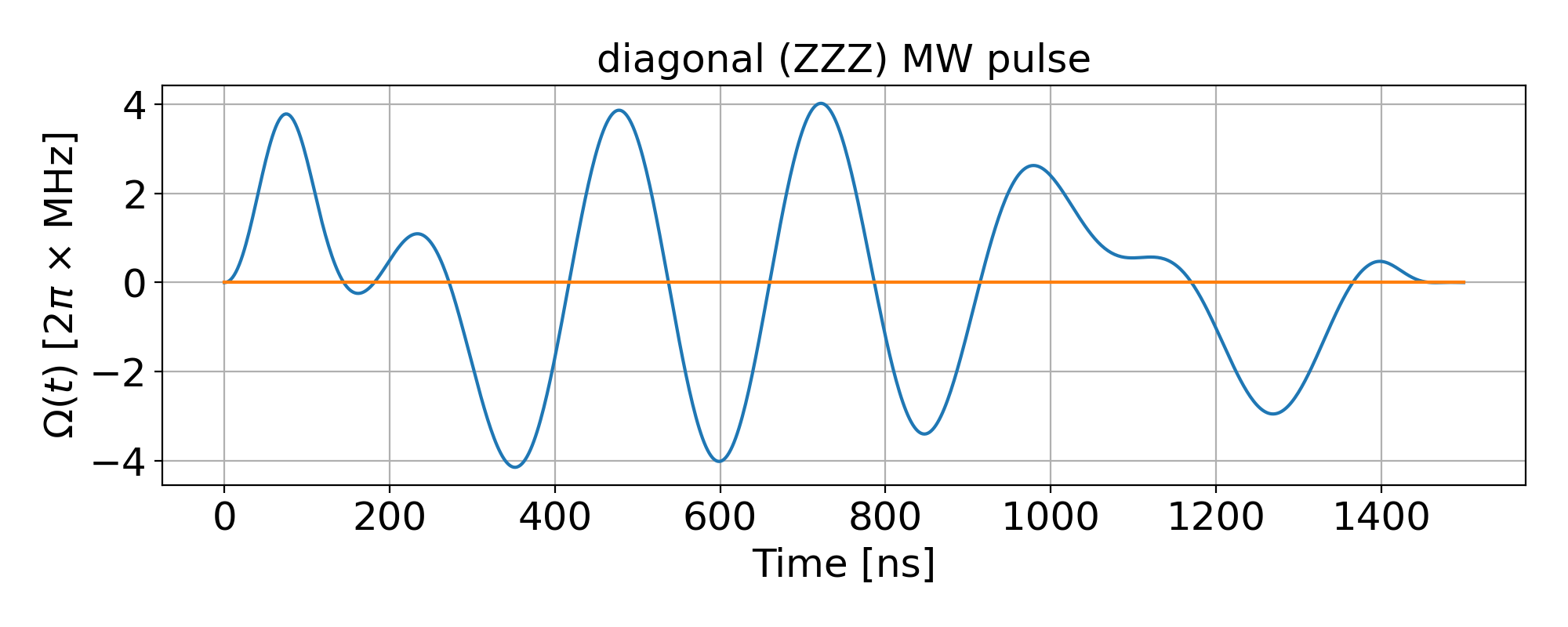}
\caption{
Optimized time-modulated Rabi-frequency $\Omega(t)$ implementing a three-body interaction phase $\phi(\{a,b,c\})\approx\pi/4$ (modulo $\pi$), while suppressing all pairwise interaction phases $\phi(\{a,b\})$, $\phi(\{a,c\})$, and $\phi(\{b,c\})$, obtained via phase-invariant optimization.
}
\label{fig:nv_3q_pulse_ZZZ}
\end{figure}

The resulting optimized microwave pulse is shown in Fig.~\ref{fig:nv_3q_pulse_ZZZ}. The pulse duration is $T=1500$~ns, and the parametrization consists of $N_c = 8$ frequency terms, as given in Eq.~\eqref{eq:pulse_parametrization}. The used parameters are provided in the public repository~\cite{baran2026_multiqubit_control} in the form of a parameter vector as in Eq.~\eqref{eq:pulse_parametrization_example}. It exhibits smooth amplitude and phase modulation while realizing the prescribed interaction structure with a single shaped pulse. 

\begin{figure}[h!]
\centering
\includegraphics[width=0.49\textwidth]{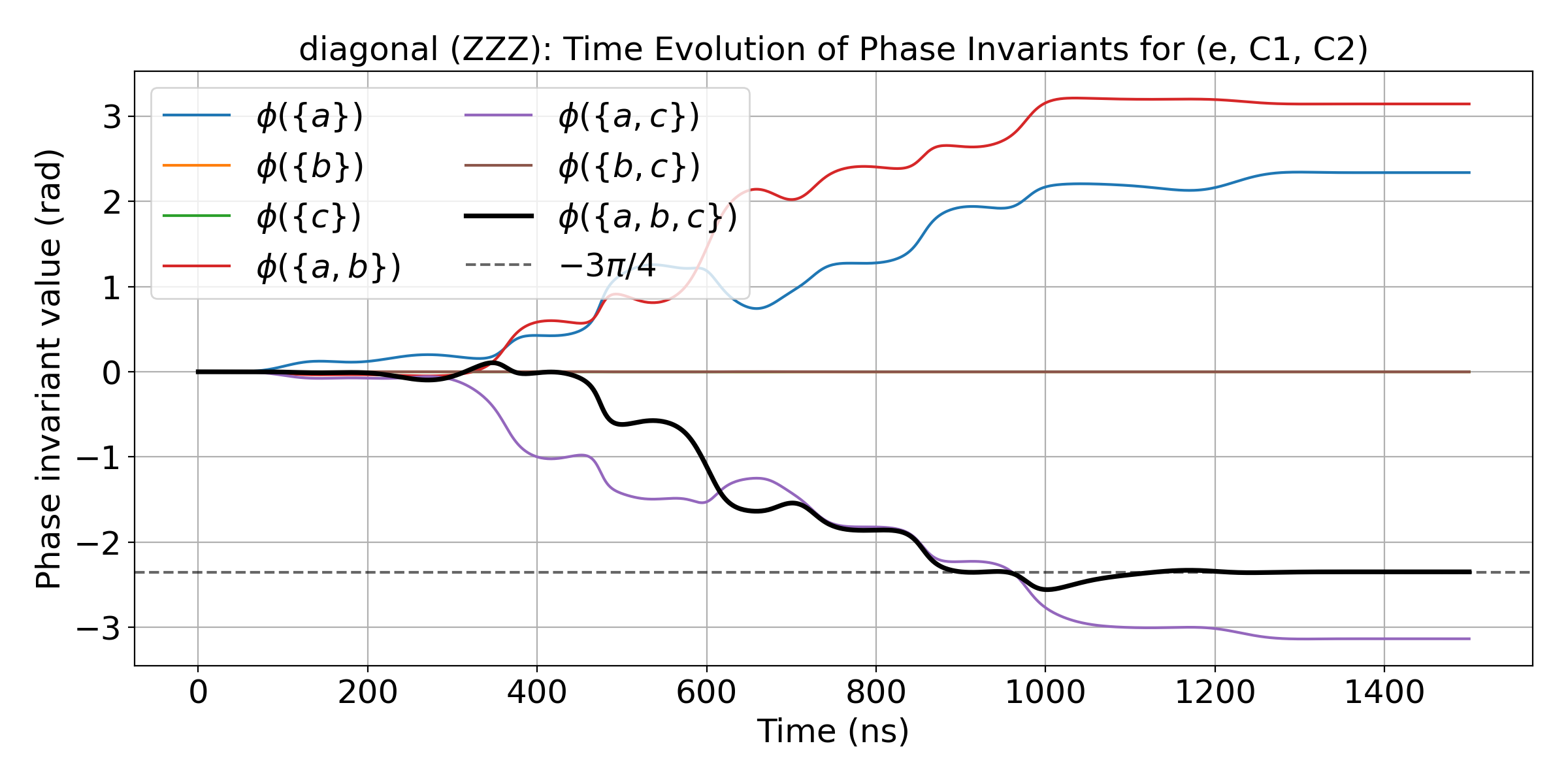}
\caption{
Time evolution of the phase invariants $\phi_t(S)$ during the optimized pulse. The three-body invariant $\phi_t(\{a,b,c\})$ (black curve) approaches $-3\pi/4$, which is equivalent to $\pi/4$ modulo $\pi$. Pairwise invariants approach integer multiples of $\pi$, indicating vanishing two-body interaction contributions, given the $\pi$-periodicity~\eqref{eq:pi_peridicity}.
}
\label{fig:nv_3q_invariants_vs_time_ZZZ}
\end{figure}

To validate the interaction dynamics we evaluate the $k$-body phase decomposition  throughout the evolution by tracking the time-dependent phase invariants $\phi_t(S)$. This reveals how the control pulse builds up the desired three-body interaction phase $\phi_t(\{a,b,c\})$ while guiding the pairwise phase components $\phi_t(\{a,b\})$, $\phi_t(\{a,c\})$, and $\phi_t(\{b,c\})$ to be suppressed towards phase equal to zero, up to the inherent $\pi$-periodicity~\eqref{eq:pi_peridicity}. As shown in Fig.~\ref{fig:nv_3q_invariants_vs_time_ZZZ}, the three-body invariant $\phi_t(\{a,b,c\})$ builds up to the target value while the pairwise invariants end up suppressed, up to the inherent $2\pi$-periodicity of the phase map. 
\\

\noindent
Residual single-qubit phases can be removed via virtual $Z$ corrections determined from the first-order phase invariants,
\begin{equation}
U_{\mathrm{loc}}
=
\exp\!\Bigl(-i\sum_{j\in\{a,b,c\}}\phi(\{j\})\, Z_j\Bigr).
\end{equation}
The resulting operation
\begin{equation}
U_{\mathrm{final}} = U_{\mathrm{loc}}\,U^{A\otimes B\otimes C}_{(\mathrm{pulse})}(\mathbf p,T)
\end{equation}
achieves a fidelity
\begin{equation}
F_{\mathrm{ZZZ}}
=
\frac{\big|\mathrm{Tr}\!\left[
(U_{\mathrm{target}})^{\dagger}U_{\mathrm{final}}
\right]\big|^{2}}{8^{2}}
= 0.9978 .
\end{equation}

\noindent
This example illustrates how phase invariants provide direct access to the interaction structure of a quantum operation. By formulating the control objective in terms of the support selective phase invariants, prescribed many-body interaction contribution terms can be synthesized while unwanted terms are suppressed. The resulting pulse realizes the target interaction structure without in one control step, demonstrating how interaction-selective control objectives can be implemented in practice.

\subsection{Non-diagonal Three-Qubit Entangler}
\label{sec:results_nv_xzz}

\noindent
We now apply the phase-invariant control framework of Sec.~\ref{sec:control_objective_phase_invariants} to synthesize a non-diagonal three-qubit entangling gate by performing the optimization in a diagonalizing frame. The target operation is
\begin{equation}
\label{eq:desired_u_nondiag}
U_{\mathrm{target}} = e^{\,i\frac{\pi}{4}X\!\otimes\!Z\!\otimes\!Z},
\end{equation}
which also generates a GHZ state up to local phases and thus produces genuine tripartite entanglement~\cite{coffman2000distributed,dur2000three}. This target gate is not naturally diagonal and therefore serves as an explicit example of the framework in a diagonalizing frame for arbitrary gates. \\


\noindent
As before, we simulate a solid-state register consisting of an NV electronic spin ($A$) coupled to two proximal $^{13}$C nuclear spins ($B$, $C$), with the $^{14}$N nuclear spin fixed in a spectator $m_I$ manifold. The system dynamics are described by the effective Hamiltonian in Eq.~\eqref{eq:H_NV_RWA_main}, and the microwave control fields are parameterized according to Eq.~\eqref{eq:pulse_parametrization_example}. We therefore again consider the projected propagator $U^{A\otimes B\otimes C}_{(\mathrm{pulse})}(\mathbf p,T)$, corresponding to the restriction of the full propagator $U_{\mathrm{pulse}}(\mathbf p,T)$ defined in Eq.~\eqref{eq:U_pulse_definition} to the logical $A\!\otimes\!B\!\otimes\!C$ subspace.\\

\noindent
The phase invariants for this target gate are constructed in the \emph{Hadamard-sandwiched frame},
\begin{equation}
\label{eq:frame_transformation}
\begin{aligned}
U_{\mathrm{diag}} &= (H_A\!\otimes\!I_B\!\otimes\!I_C)\,
U^{A\otimes B\otimes C}_{(\mathrm{pulse})}(\mathbf p,T)\,
(H_A\!\otimes\!I_B\!\otimes\!I_C) \\
\end{aligned}
\end{equation}
where the target unitary is diagonal and so the the propagator can be analyzed in the computational basis phases directly by constructing phase-invariant, as given in Secs.~\ref{sec:interaction_resolved_representation}–\ref{sec:control_objective_phase_invariants}. After optimization, the Hadamard gates can be removed such that the same physical pulse implements the desired non-diagonal action.

\medskip
\noindent
In the Hadamard-sandwiched frame~\eqref{eq:frame_transformation}, the optimization proceeds as for the previous target gate, using the phase invariants introduced in Sec.~\ref{sec:interaction_resolved_representation}, with the control objective being by specifying the target interaction phases
\begin{equation}
\begin{aligned}
&\phi^\star(\{a,b,c\}) = \tfrac{\pi}{4}, \\
&\phi^\star(\{a,b\})
=
\phi^\star(\{a,c\})
=
\phi^\star(\{b,c\})
= 0,
\end{aligned}
\label{eq:Uxzzinvariants}
\end{equation}
with the optimization minimizing the $\pi$-periodic cost function $\mathcal{J}_{\mathrm{3q}}$ defined in Eq.~\eqref{eq:phase-invariant-cost},  while the weights are chosen as: $w(\{{a,b,c}\})=0.5$, $w(|S|=2)=0.2$, $w(|S|=1)=0.0$. Explicit formulas for the used phase invariants in terms of the computational-basis phase values $\{\phi_x\}_{x\in\{0,1\}^3}$ are given in Appendix~\ref{app:explicit_3q_invariants}. Using these phase invariants we optimize a control pulse which steers the system toward the same prescribed many-body interaction structure in Eq.~\eqref{eq:Uzzzinvariants} as it is equivalent in the hadamard-sandwiched frame in Eq.~\eqref{eq:frame_transformation}.

\medskip
\noindent
After optimization, we return to the computational basis by discarding the Hadamard gates in Eq.~\eqref{eq:frame_transformation}, such that the effective action of the pulse is
\begin{equation}
U^{A\otimes B\otimes C}_{(\mathrm{pulse})}(\mathbf p,T)
\;\approx\;
e^{i\frac{\pi}{4}X\!\otimes\!Z\!\otimes\!Z}
\end{equation}
up to local transformations.

\begin{figure}[h!]
\centering
\includegraphics[width=0.47\textwidth]{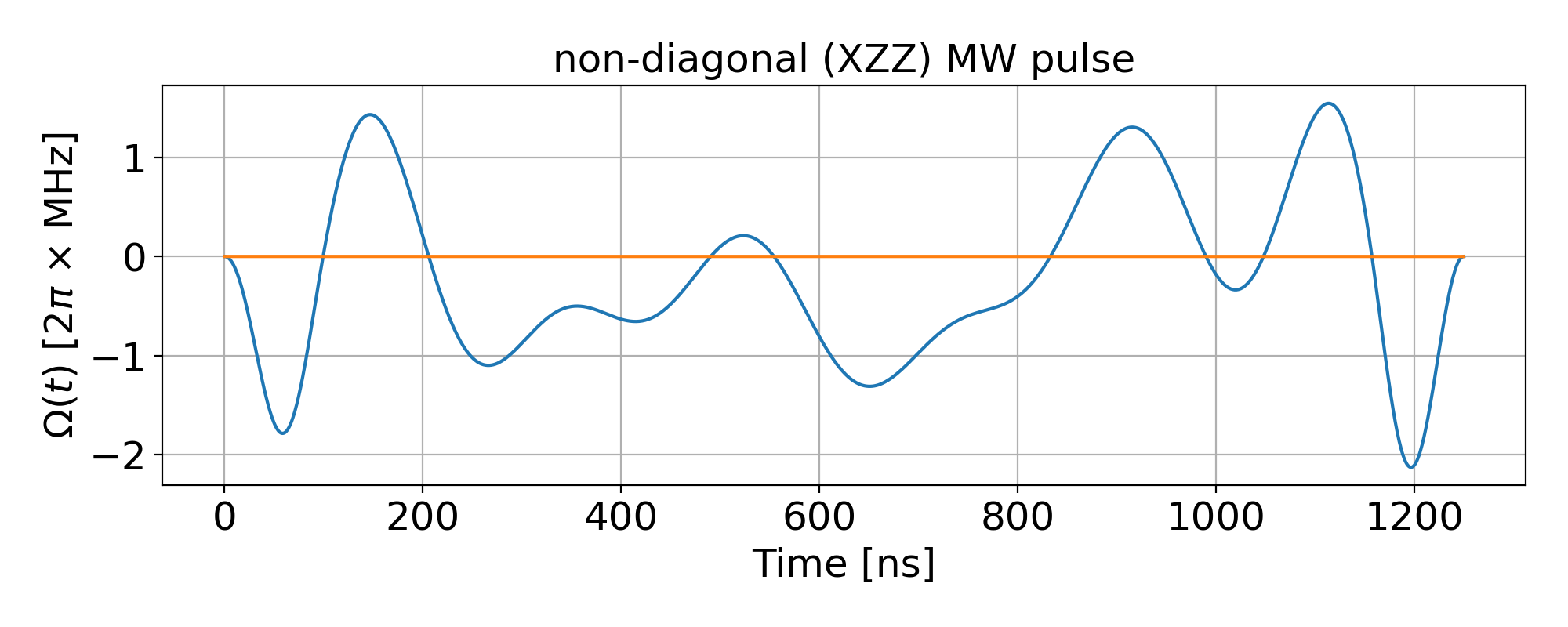}
\caption{
Optimized time-modulated Rabi-frequency $\Omega(t)$ implementing the non-diagonal three-qubit entangler $e^{i(\pi/4)X\!\otimes\!Z\!\otimes\!Z}$ in the computational basis, synthesized via phase-invariant optimization in the diagonalizing (Hadamard) frame.
}
\label{fig:nv_3q_pulse_xzz}
\end{figure}

\medskip
\noindent
The optimized microwave envelope (Fig.~\ref{fig:nv_3q_pulse_xzz}) exhibits smooth amplitude and phase modulation over $T=1250$~ns, and consists of $N_c = 11$ frequency terms, as given in Eq.~\eqref{eq:pulse_parametrization}. The used parameters are provided in the public repository~\cite{baran2026_multiqubit_control} in the form of a parameter vector as in Eq.~\eqref{eq:pulse_parametrization_example}.  It realizes the control objective in the diagonalizing (Hadamard) frame as a single shaped pulse. After the removal of the diagonalizing transformations, the same physical pulse implements the target non-diagonal entangler in the computational basis.

\begin{figure}[h!]
\centering
\includegraphics[width=0.49\textwidth]{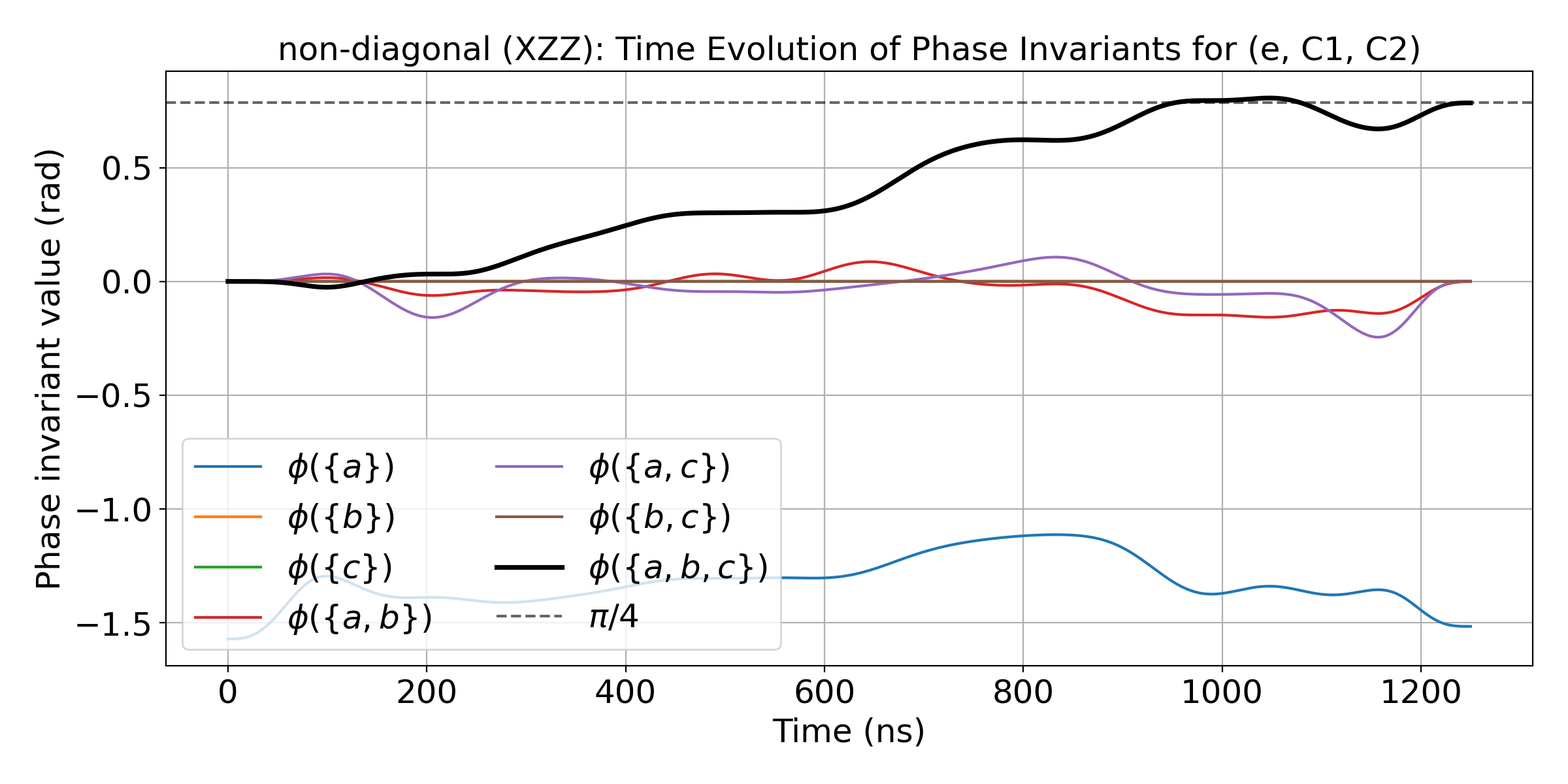}
\caption{
Time evolution of the phase invariants $\phi_t(S)$ extracted from the phase map in the diagonalizing (Hadamard-sandwiched) frame during the optimized non-diagonal gate pulse. As in the canonically diagonal case, in Fig.~\ref{fig:nv_3q_invariants_vs_time_ZZZ}, the three-body phase invariant $\phi_t(\{a,b,c\})$ (thick black line) converges to $\pi/4$, realizing the desired three-body interaction. The two-body phase invariants $\phi_t(\{a,b\})$, $\phi_t(\{a,c\})$, and $\phi_t(\{b,c\})$ directly go to zero in this case, corresponding to vanishing pairwise interaction content.
}
\label{fig:nv_3q_invariants_vs_time_xzz}
\end{figure}

\noindent
Next we validate the interaction dynamics by tracking the time evolution of the phase invariants $\phi_t(S)$ for all nonempty subsets $S\subseteq\{a,b,c\}$. This reveals that the control pulse builds up the desired three-body interaction phase $\phi_t(\{a,b,c\})$ in the diagonalizing frame while guiding the pairwise phase components $\phi_t(\{a,b\})$, $\phi_t(\{a,c\})$, and $\phi_t(\{b,c\})$ to be suppressed toward zero, up to the inherent $2\pi$-periodicity~\eqref{eq:pi_peridicity}. Figure~\ref{fig:nv_3q_invariants_vs_time_xzz} shows the resulting phase-invariant trajectories.

\medskip
\noindent
Residual local phases can be assessed and corrected via the first-order phase invariants in the diagonalizing frame and yield the local correction operator
\begin{equation}
U_{\mathrm{loc}}^{(\mathrm{diag})}
=
\exp\!\Bigl(-i\sum_{j\in\{a,b,c\}}\phi(\{j\})\, Z_j\Bigr).
\end{equation}
Transforming back to the computational basis gives
\begin{equation}
\begin{aligned}
U_{\mathrm{loc}}^{(\mathrm{comp})}
&=
(H_A\!\otimes\!I_B\!\otimes\!I_C)\,
U_{\mathrm{loc}}^{(\mathrm{diag})}\,
(H_A\!\otimes\!I_B\!\otimes\!I_C) \\
&=
e^{i(\phi(\{a\}) X_A + \phi(\{b\}) Z_B + \phi(\{c\}) Z_C)} .
\end{aligned}
\end{equation}
Note that here the local $X$ rotation cannot be corrected virtually and an additional single-qubit gate is needed. The final operation is then
\begin{equation}
U_{\mathrm{final}}
=
U_{\mathrm{loc}}^{(\mathrm{comp})}\,U^{A\otimes B\otimes C}_{(\mathrm{pulse})}(\mathbf p,T).
\end{equation}

\medskip
\noindent
The fidelity with the target operation in Eq.~\eqref{eq:desired_u_nondiag} evaluates to
\begin{equation}
F_{\mathrm{XZZ}}
=
\frac{\big|\mathrm{Tr}\!\left[
\left(U_{\mathrm{target}}\right)^{\!\dagger}
U_{\mathrm{final}}
\right]\big|^{2}}{8^{2}}
= 0.9985,
\end{equation}
demonstrating that interaction-resolved optimization in a diagonalizing frame extends naturally to high-fidelity non-diagonal multiqubit gates.

\medskip
\noindent
This example demonstrates that the interaction-resolved control objectives apply also to non-diagonal multi-qubit gates. By performing the optimization in a suitable diagonalizing frame, the interaction structure of an arbitrary Pauli operator can be accessed through the same phase invariants. 

\section{Conclusion and Outlook}
\label{sec:conclusion}

\noindent
We introduced an interaction-resolved representation for multi-qubit unitaries based on the phases acquired by computational-basis states in a suitable diagonalizing frame. In this representation the interaction structure of a unitary transformation is encoded in the phases associated with computational-basis configurations. By forming parity-weighted sums of these phases using the eigenvalue structure of Pauli-$Z$ strings, we obtain quantities that resolve the many-body interaction contributions supported on arbitrary subsets of qubits, which we refer to as \emph{support-selective phase invariants}. Each invariant isolates the many-body interaction contribution associated with a specific subset while canceling contributions supported on all other subsets. The resulting phase invariants provide an interaction-resolved coordinate system that gives direct access to local, pairwise, tripartite, and general \(k\)-partite interaction content underlying multipartite entangling operations. This naturally enables the formulation of interaction-selective control objectives directly in terms of a desired many-body interaction structure.\\

\noindent
To demonstrate the utility of this decomposition, we presented two, numerical, quantum optimal control examples for a representative solid state platform, a realistic simulated NV-$^{14}$N-$^{13}$C-$^{13}$C register. First, we considered a diagonal three-qubit entangler, showing how phase-invariant control objectives directly specify and synthesize the desired many-body interaction structure of a unitary transformation. In this case the tripartite interaction $e^{i(\pi/4)Z\otimes Z\otimes Z}$ was synthesized using a single shaped microwave pulse with fidelity $F_{\mathrm{ZZZ}}=0.9978$. We then considered a non-diagonal three-qubit interaction of the form $e^{i(\pi/4)X\otimes Z\otimes Z}$, obtained by performing the optimization in a suitable diagonalizing frame. This second example demonstrates that the framework extends naturally non-diagonal Pauli interactions: by transforming to a frame in which the relevant interaction becomes diagonal, the same interaction-resolved control objectives can be applied. The resulting operation was realized with fidelity $F_{\mathrm{XZZ}}=0.9985$. The gate durations, of $1.25\,\mu\mathrm{s}$ and $1.5\,\mu\mathrm{s}$, remain well below the coherence time~\cite{hanson2008coherent,naydenov2011dynamical,balasubramanian2009ultralong,maurer2012room,childress2006coherent,taminiau2014universal}. In both cases the prescribed  interaction structure was synthesized through support selective phase invariant coordinate control objectives, while leaving local, easily correctable phases unconstrained. These results show that nontrivial higher-weight multi-qubit interactions can be synthesized by a single pulse. Many stabilizer measurements in quantum error correction are typically decomposed into sequences of two-qubit gates and ancilla-mediated operations~\cite{gottesman1997stabilizer,fowler2012surface,litinski2019game,taminiau2014universal}. Our results demonstrate that isolated tripartite entangling operations can be synthesized  as single shaped pulses in the considered platform model  thereby suggesting a route toward reduced circuit depth and thereby potentially lower accumulated control errors. 

\noindent
Looking ahead, promising directions include experimental realization on multi-spin NV registers and related AMO platforms, as well as closed-loop interaction-resolved characterization and optimization based on support-selective phase invariants. In this context, the phase map underlying the invariants can be reconstructed experimentally via conditional Ramsey interferometry, as described explicitly for the three-qubit case in Appendix~\ref{app:explicit_3q_phase_map_reconstruction}. These measurements provide direct access to relative phases between computational-basis states and could therefore serve as observables in a closed-loop control protocol, enabling interaction-resolved characterization and therefore optimization without requiring full process tomography. 


\bibliographystyle{unsrt}  
\bibliography{references}

\appendix

\section*{Acknowledgements}
\noindent
The authors acknowledge fruitful discussions with Thomas Reisser, Marcus Doherty and Robert Zeier. This work was supported by the European Union’s HORIZON Europe program via project SPINUS (No. 101135699), project QCFD (101080085, HORIZON-CL4-2021 DIGITAL-EMERGING02-10), project OpenSuperQ-Plus100 (101113946, HORIZON-CL4-2022-QUANTUM-01-SGA), by AIDAS-AI, Data Analytics and Scalable Simulation, which is a Joint Virtual Laboratory gathering the Forschungszentrum Juelich and the French Alternative Energies and Atomic Energy Commission, and by Germany’s Excellence Strategy - the Cluster of Excellence Matter and Light for Quantum Computing (ML4Q2) EXC 2004/2–390534769.\\

\noindent
The authors declare no competing financial interest.

\section*{Author Contributions}
\noindent
The project originated from discussions among all authors on how to formulate optimal control objectives that selectively target multi-qubit interaction terms without requiring full-unitary specification. B.B. derived the computational-basis phase based $k$-body interaction decomposition used in this work and formulated its integration into an optimal control framework, with input from M.M.M. and F.M.. M.M.M. and F.M. supervised the project. B.B. implemented the method numerically. T.C., M.M.M. and F.M. provided scientific guidance and critical analysis throughout its development. B.B. drafted the manuscript. All authors contributed to, reviewed and approved the final manuscript.



\section*{Data Availability}
\noindent
The data and reproduction code is openly available on Zenodo and GitHub \cite{baran2026_multiqubit_control}.


\section{Full Derivation of the NV-Center Hamiltonian}
\label{app:nv_model}

This appendix details the derivation of the effective NV-center Hamiltonian given in Sec.\ref{sec:spin_model}, inspired by the perturbative analysis of Chen \emph{et al.}~\cite{Chen}.

\paragraph{Laboratory Frame.}
Throughout, Hamiltonian parameters are expressed in angular-frequency units. With a static magnetic field $B_0$ aligned to the NV axis and a linearly polarized microwave (MW) field $B_{1,\mathrm{MW}}(t)$ applied orthogonally, the coupled NV-nuclear system is described by
\begin{equation}
\begin{aligned}
H_{\mathrm{NV}}
&= D\!\left(S_z^2-\tfrac{2}{3}\right)
+ \gamma_e B_0 S_z
+ \gamma_e B_{1,\mathrm{MW}}(t)\,S_x\\
&\hspace{-0.7cm}+ \sum_i \vec{S}\!\cdot\! A_i \!\cdot\! \vec{I}_i 
\quad - \sum_i \gamma_i B_0 I_{iz}
\;+\; \sum_i Q_i\!\left(I_{iz}^2-\tfrac{2}{3}\right),
\end{aligned}
\label{eq:H_full}
\end{equation}
where $D/2\pi=2.87\,\mathrm{GHz}$ is the zero-field splitting, $\vec S$ ($S=1$) are the NV electronic spin operators, $\vec I_i$ are nuclear spins ($^{14}$N, selected $^{13}$C) with gyromagnetic ratios $\gamma_i$, $A_i$ are hyperfine tensors, and $Q_i$ are nuclear quadrupole terms.\\

\paragraph{Two-level reduction and secular approximation.}
We consider the effect of applying a strong magnetic field $B_0$. Together with the zero-field splitting $D$, a strong magnetic field lifts the degeneracy of the $m_s\pm1$ electronic levels and isolates the $\{\ket{0_e},\ket{-1_e}\}$ transition from the remaining electronic level $\ket{+1_e}$, enough for it to be neglected (valid for $B_0\gtrsim 0.5\,$T). The electronic spin can thereby be treated as an effective two-level transition between $\{\ket{0_e},\ket{-1_e}\}$, with transition frequency
\[
\Lambda_s = D - \gamma_e B_0.
\]
With that, we write the electronic Hamiltonian as
\begin{equation}
H_e = \frac{\Lambda_s}{2}\,\sigma_z \;+\; \gamma_e B_{1,\mathrm{MW}}(t)\,\frac{\sigma_x}{\sqrt{2}}.
\label{eq:He_def}
\end{equation}

Given that the electronic splitting $\Lambda_s$ is much larger than the hyperfine couplings $A_i$, the terms in the hyperfine interaction \(\sum_i \vec S\cdot A_i\cdot \vec I_i\) that are non-diagonal in the electronic basis, average out on relevant timescales. This motivates a so-called secular-approximation: only the electron-diagonal components of the hyperfine interaction are retained, so that
\[
\sum_i \vec S\cdot A_i\cdot \vec I_i \longrightarrow
S_z\left(A_{i,zz}I_{iz}+
A_{i,zx}I_{ix}+
A_{i,zy}I_{iy}\right)
\]
Further, we have to consider that in the $\ket{-1_e}$-manifold, the projected hyperfine interaction contributes to the nuclear dynamics with negative sign, giving
\begin{equation}
\begin{aligned}
H_{-1} &= -\sum_i \gamma_i B_0 I_{iz} \;+\; \sum_i Q_i\!\left(I_{iz}^2-\tfrac{2}{3}\right)\\
&\hspace{0.43cm}\;-\; \sum_i \big(A_{i,zz} I_{iz}+A_{i,zx} I_{ix}+A_{i,zy} I_{iy}\big) \label{eq:H-1_def}
\end{aligned}
\end{equation}
By contrast, in the $\ket{0_e}$-manifold the projected $S_z$ contribution vanishes, since $S_z\ket{0_e}=0$, so that the nuclei evolve only under their Zeeman $\gamma_i B_0 I_{iz}$ and quadrupolar terms $Q_i\!\left(I_{iz}^2-\tfrac{2}{3}\right)$,
\begin{equation}
\begin{aligned}
H_{0} &= -\sum_i \gamma_i B_0 I_{iz} \;+\; \sum_i Q_i\!\left(I_{iz}^2-\tfrac{2}{3}\right)
\label{eq:H0_def}
\end{aligned}
\end{equation}
Recombining the secular electronic Hamiltonian with the two conditional nuclear Hamiltonians yields the block form
\begin{equation}
H_{\mathrm{NV}}
= H_e
+ \ket{0_e}\!\bra{0_e}\otimes H_{0}
+ \ket{-1_e}\!\bra{-1_e}\otimes H_{-1}.
\label{eq:block_form}
\end{equation}

\paragraph{Diagonalizing the Nuclear Basis.}
We want to diagonalize the $H_{-1}$ Hamiltonian because it contains off-diagonal terms with respect to the electronic quantization axis, in contrast to the $H_{0}$ Hamiltonian.
In the $H_{-1}$ Hamiltonian, the transverse hyperfine components tilt the effective nuclear field away from the NV axis. For nucleus $i$, the combined secular hyperfine and nuclear Zeeman contributions in the $H_{-1}$ Hamiltonian can be grouped as
\begin{equation}
    H^{(i)}_{-1,\mathrm{hf}} = (A_{i,zz}+\gamma_iB_0)I_{iz}
    +A_{i,zx}I_{ix}
    ++A_{i,zy}I_{iy}
    \label{eq:Hminus_hf}
\end{equation}
which, by itself, can be geometrically interpreted as an interaction of the nuclear spin with an effective static field
\[
\vec b_i = 
\begin{pmatrix}
A_{i,zx}\\
A_{i,zy}\\
A_{i,zz}+\gamma_iB_0\\
\end{pmatrix}
\]
whose orientation corresponds to the current quantization axis in the $H_{-1}$ Hamiltonian.

In order to ultimately rotate the basis of $H_{-1}$, we parameterize the orientation of $\vec b_i$ by spherical angles $\theta_i,\phi_i$ \cite{schweiger2001principles},
\begin{equation}
\tan\phi_i = \frac{A_{i,zy}}{A_{i,zx}},
\qquad
\tan\theta_i = \frac{\sqrt{A_{i,zx}^2+A_{i,zy}^2}}
{A_{i,zz}+\gamma_iB_0},
\label{eq:angles}
\end{equation}
and define its magnitude
\begin{equation}
\omega_i = |\vec b_i| =
\sqrt{ (A_{i,zz}+\gamma_iB_0)^2
+ A_{i,zx}^2
+ A_{i,zy}^2 }.
\label{eq:omega_i}
\end{equation}

Based on these spherical angles, we introduce a rotated nuclear basis $(x'_i,y'_i,z'_i)$ whose quantization axis $z'_i$ is aligned with $\vec b_i$. The $I_{iz}$ operator in this rotated basis basis becomes 
\begin{equation}
    I'_{iz} = I_{ix}\sin\theta_i\cos\phi_i 
    + I_{iy}\sin\theta_i\sin\phi_i 
    + I_{iz}\cos\theta_i.
    \label{eq:Izprime}
\end{equation}
With this choice, the transverse contribution in the $H_{-1}$ Hamiltonian become diagonal:
\[
H^{(i)}_{-1,\mathrm{hf}}= \vec b_i\cdot\vec I_i = \omega_i I'_{iz}.
\]

The $H_{-1}$ Hamiltonian therefore takes the form
\begin{align}
H_{-1}
&=
-\sum_i\omega_i I'_{iz}
+
\sum_iQ_i\!\left(I_{iz}'^{\,2}-\tfrac23\right).
\label{eq:H-1_rot}
\end{align}

Thus, after rotating the nuclear basis, the quantization axes associated with the $H_{-1}$ Hamiltonian are collected onto the effective hyperfine axis of each nucleus, rendering the Hamiltonian diagonal in the rotated basis.\\

\paragraph{Electronic transitions between nuclear quantization frames.}

While the $H_0$ Hamiltonian remains quantized along the NV axis, the $H_{-1}$ Hamiltonian is now quantized along the effective hyperfine axis defined by $\vec b_i$. The two nuclear Hamiltonians therefore have a relative tilt between their quantization frames.

The rotated operator \(I'_{iz}\) implicitly defines a nuclear rotation
\(R_i\) satisfying
\[
I'_{iz}=R_i I_{iz} R_i^\dagger.
\]
Since the effective field \(\vec b_i\) is parameterized by the spherical angles \((\theta_i,\phi_i)\), this rotation is generated by
\[
R_i
=
\exp\!\left[
-i\theta_i
\left(
-I_{ix}\sin\phi_i
+
I_{iy}\cos\phi_i
\right)
\right],
\]
with collective rotation
\[
R=\prod_iR_i.
\]

For the weakly misaligned nuclei considered here,  \(\theta_i\ll1\), the rotation may be expanded to first order,
\[
R
\simeq
1-iG,
\]
where
\[
G
=
\sum_i\theta_i
\left(
-I_{ix}\sin\phi_i
+
I_{iy}\cos\phi_i
\right).
\]

Rather than rewriting both nuclear Hamiltonians in a common operator basis, we absorb this relative nuclear rotation into the electronic transition operators. Defining the electronic raising and lowering operators as
\[
\sigma_+
=
\ket{0_e}\!\bra{-1_e},
\qquad
\sigma_-
=
\ket{-1_e}\!\bra{0_e},
\]
the microwave interaction
\[
H_{\mathrm{MW}}
=
\frac{\gamma_eB_{1,\mathrm{MW}}(t)}{\sqrt2}
(\sigma_+ + \sigma_-)
\]
therefore becomes
\begin{equation}
\begin{aligned}
H_{\mathrm{MW}}
&=
\frac{\gamma_eB_{1,\mathrm{MW}}(t)}{\sqrt2}
\left(
\sigma_+\otimes R^\dagger
+
\sigma_-\otimes R
\right)
\\
&\approx
\frac{\gamma_eB_{1,\mathrm{MW}}(t)}{\sqrt2}
\left[
\sigma_x
-
\sigma_y G
\right].
\end{aligned}
\label{eq:HMW_rot}
\end{equation}

The first term describes the bare electronic microwave transition, while the second term is a first-order electron-nuclear modulation arising from the mismatch between the nuclear quantization frames of the two electronic manifolds.\\

\paragraph{Interaction picture.}

The next step of the derivation is a transformation to the interaction picture that removes the free precession dynamics and leaves only the control-induced dynamics explicit. We therefore decompose the Hamiltonian as
\begin{equation}
H_{\mathrm{NV}}
=
H_{\mathrm{free}}
+
H_{\mathrm{MW}},
\label{eq:H_split}
\end{equation}
where the free evolution is generated by the diagonal electronic and nuclear
terms,
\begin{equation}
\begin{aligned}
H_{\mathrm{free}}
&=
\frac{\Lambda_s}{2}\sigma_z
\\
&\quad
+
\ket{0_e}\!\bra{0_e}\otimes
\left[
-\sum_i\gamma_iB_0 I_{iz}
+
\sum_iQ_i\!\left(I_{iz}^{\,2}-\tfrac23\right)
\right]
\\
&\quad
+
\ket{-1_e}\!\bra{-1_e}\otimes
\left[
-\sum_i\omega_i I'_{iz}
+
\sum_iQ_i\!\left(I_{iz}'^{\,2}-\tfrac23\right)
\right],
\end{aligned}
\label{eq:H_free}
\end{equation}
while the microwave interaction is
\begin{equation}
\begin{aligned}
H_{\mathrm{MW}}
&\approx
\frac{\gamma_e B_{1,\mathrm{MW}}(t)}{\sqrt2}
\left[
\sigma_x
-
\sigma_y G
\right]\,.
\end{aligned}
\label{eq:HMW_rot}
\end{equation}

We choose \(H_{\mathrm{free}}\) as the generator of the interaction picture, such that
\begin{equation}
H_{\mathrm{NV}}^{(I)}(t)
=
e^{iH_{\mathrm{free}}t}
H_{\mathrm{MW}}
e^{-iH_{\mathrm{free}}t}.
\label{eq:IP_transform}
\end{equation}

We first transform the bare electronic part of the microwave interaction,
\[
H_{\mathrm{MW},x}
=
\frac{\gamma_e B_{1,\mathrm{MW}}(t)}{\sqrt2}\sigma_x .
\]
Its interaction-picture form is
\begin{equation}
H_{\mathrm{MW},x}^{(I)}(t)
=
\frac{\gamma_e B_{1,\mathrm{MW}}(t)}{\sqrt2}
e^{iH_{\mathrm{free}}t}
\sigma_x
e^{-iH_{\mathrm{free}}t}.
\label{eq:HMWx_IP_start}
\end{equation}

Since \(H_{\mathrm{free}}\) is diagonal in the electronic basis and in the natural nuclear basis of each manifold, its propagator acts only through energy-dependent phase accumulation,
\[
e^{iH_{\mathrm{free}}t}
=
\sum_{e,\mathbf m}
e^{iE_e(\mathbf m)t}
\ket{e,\mathbf m}\!\bra{e,\mathbf m}.
\]
Using \(\sigma_x=\sigma_++\sigma_-\), the interaction picture evolution of the electronic transition operator is therefore governed by the energy difference between the connected free eigenstates,
\begin{equation}
\begin{aligned}
&e^{iH_{\mathrm{free}}t}
\sigma_x
e^{-iH_{\mathrm{free}}t}
\notag\\
&=
\sum_{\mathbf m}
\Big[
\sigma_+
e^{i[E_0(\mathbf m)-E_{-1}(\mathbf m)+\Lambda_s]t}
\\
&\hspace{1.5cm}
+
\sigma_-
e^{-i[E_0(\mathbf m)-E_{-1}(\mathbf m)+\Lambda_s]t}
\Big]
\otimes
\ket{\mathbf m}\!\bra{\mathbf m}.
\end{aligned}
\label{eq:sigmax_IP_phase}
\end{equation}
Defining the configuration-dependent transition frequency
\begin{equation}
\begin{aligned}
\Lambda(\mathbf m)
&:=
\Lambda_s
+
E_0(\mathbf m)
-
E_{-1}(\mathbf m)
\\
&=
\Lambda_s
+
\left[
-\sum_i\gamma_iB_0m_i
+
\sum_iQ_i\!\left(m_i^2-\tfrac23\right)
\right]
\\
&\quad
-
\left[
-\sum_i\omega_i m_i
+
\sum_iQ_i\!\left(m_i^2-\tfrac23\right)
\right]
\\
&=
\Lambda_s
-
\sum_i m_i(\gamma_iB_0-\omega_i).
\end{aligned}
\label{eq:Lambda_config_general}
\end{equation}
this becomes
\begin{equation}
\begin{aligned}
&e^{iH_{\mathrm{free}}t}
\sigma_x
e^{-iH_{\mathrm{free}}t}
\notag\\
&=
\sum_{\mathbf m}
\left[
\sigma_+ e^{i\Lambda(\mathbf m)t}
+
\sigma_- e^{-i\Lambda(\mathbf m)t}
\right]
\otimes
\ket{\mathbf m}\!\bra{\mathbf m}
\\
&=
\sum_{\mathbf m}
\big[
\sigma_x\cos(\Lambda(\mathbf m)t)
-
\sigma_y\sin(\Lambda(\mathbf m)t)
\big]
\otimes
\ket{\mathbf m}\!\bra{\mathbf m}.
\end{aligned}
\label{eq:sigmax_IP_step}
\end{equation}

Consequently, the bare electronic part of the microwave interaction becomes
\begin{equation}
\begin{aligned}
H_{\mathrm{MW}}^{(I)}(t)
&=
\frac{\gamma_e B_{1,\mathrm{MW}}(t)}{\sqrt2}
\sum_{\mathbf m}
\big[
\sigma_x\cos(\Lambda(\mathbf m)t)
-
\sigma_y\sin(\Lambda(\mathbf m)t)
\big]
\\
&\hspace{1cm}\otimes
\ket{\mathbf m}\!\bra{\mathbf m}.
\end{aligned}
\label{eq:HMW_I}
\end{equation}
Here, the nuclear configuration is preserved to leading order during the bare
microwave-driven electronic transition.\\

We now transform the second term in the dressed microwave interaction,
\[
H_{\mathrm{MW},eN}
=
-\frac{\gamma_e B_{1,\mathrm{MW}}(t)}{\sqrt2}
\sigma_y\otimes G .
\]
Its interaction-picture form is
\begin{equation}
H_{\mathrm{MW},eN}^{(I)}(t)
=
-\frac{\gamma_e B_{1,\mathrm{MW}}(t)}{\sqrt2}
e^{iH_{\mathrm{free}}t}
\left(
\sigma_y\otimes G
\right)
e^{-iH_{\mathrm{free}}t}.
\label{eq:HeN_IP_start}
\end{equation}

Since the electron-nuclear modulation term contributes only at first order in the small misalignment angles \(\theta_i\), we employ a leading-order description of its interaction-picture evolution. For the electronic part, we retain the dominant bare precession at \(\Lambda_s\) rather than the full configuration-dependent transition frequency \(\Lambda(\mathbf m)\). For the nuclear part, we approximate the manifold-dependent precession by the effective mean generator
\[
H_N
=
-\sum_i \delta_i I_{iz},
\,\, \text{with}\,\,
\delta_i
=
\frac{\gamma_iB_0+\omega_i}{2}.
\]
This captures the leading averaged dynamics of the weak modulation without
tracking the full configuration-resolved evolution.

For the same reason, we also approximate the transformation as
\begin{equation}
\begin{aligned}
&e^{iH_{\mathrm{free}}t}
\left(
\sigma_y\otimes G
\right)
e^{-iH_{\mathrm{free}}t}
\notag\\
&\approx
\left(
e^{i\Lambda_s\sigma_z t/2}
\sigma_y
e^{-i\Lambda_s\sigma_z t/2}
\right)
\otimes
\left(
e^{iH_N t}
G
e^{-iH_N t}
\right).
\end{aligned}
\label{eq:sigmayG_factorized}
\end{equation}

The electronic part evaluates to
\[
e^{i\Lambda_s\sigma_z t/2}
\sigma_y
e^{-i\Lambda_s\sigma_z t/2}
=
\sigma_x\sin(\Lambda_s t)
+
\sigma_y\cos(\Lambda_s t),
\]
while the nuclear operator
\[
G
=
\sum_i\theta_i
\left(
-I_{ix}\sin\phi_i
+
I_{iy}\cos\phi_i
\right)
\]
evolves under \(H_N\) as
\[
e^{iH_N t}
G
e^{-iH_N t}
=
\sum_i\theta_i
\left[
I_{ix}\cos(\delta_i t-\phi_i)
+
I_{iy}\sin(\delta_i t-\phi_i)
\right].
\]

Therefore, to leading order,
\begin{equation}
\begin{aligned}
H_{\mathrm{MW},eN}^{(I)}(t)
&\approx
-\frac{\gamma_e B_{1,\mathrm{MW}}(t)}{\sqrt2}
\big[
\sigma_x\sin(\Lambda_s t)
+
\sigma_y\cos(\Lambda_s t)
\big]
\\
&\quad\otimes
\sum_i\theta_i
\big[
I_{ix}\cos(\delta_i t-\phi_i)
+
I_{iy}\sin(\delta_i t-\phi_i)
\big].
\end{aligned}
\label{eq:HeN_I}
\end{equation}

\paragraph{Rotating-wave approximation (RWA).}
We drive the electron at microwave frequency $\omega_{\mathrm{MW}}$ with a magnetic field
\begin{equation}
B_{1,\mathrm{MW}}(t)
=
\frac{\sqrt2\,\Omega(t)}{\gamma_e}
\cos(\omega_{\mathrm{MW}} t),
\end{equation}
where $\Omega(t)$ denotes the effective time-dependent Rabi-frequency envelope. 
Substituting this field into Eq.~\eqref{eq:HMW_I} gives the prefactor
\[
\frac{\gamma_e B_{1,\mathrm{MW}}(t)}{\sqrt2}
=
\Omega(t)\cos(\omega_{\mathrm{MW}} t).
\]
Using the product-to-sum identity $\cos\alpha\cos\beta=\tfrac12[\cos(\alpha-\beta)+\cos(\alpha+\beta)]$
and $\cos\alpha\sin\beta=\tfrac12[\sin(\beta+\alpha)+\sin(\beta-\alpha)]$,
the electron interaction-picture drive
\(
\propto \sigma_x\cos[\Lambda(\mathbf m)t]-\sigma_y\sin[\Lambda(\mathbf m)t]
\)
splits into a slow (difference-frequency) piece and explicit counter-rotating (sum-frequency) pieces:
\begin{align}
&\notag\hspace{-0.0cm}H_{e}^{(\mathrm{full})}(t)= \frac{\Omega(t)}{2}\sum_{\mathbf m}\\
&\hspace{-0.2cm}
\Big\{\underbrace{\big[\sigma_x\cos((\omega_{\mathrm{MW}}-\Lambda(\mathbf m))\,t)-\sigma_y\sin((\omega_{\mathrm{MW}}-\Lambda(\mathbf m))\,t)\big]}_{\text{slow (near-resonant)}}\nonumber\\
&\hspace{-0.2cm}
+\notag \underbrace{\big[\sigma_x\cos\big((\omega_{\mathrm{MW}}+\Lambda(\mathbf m))t\big)- 
\hspace{0.2cm} \sigma_y\sin\big((\omega_{\mathrm{MW}}+\Lambda(\mathbf m))t\big)\big]}_{\text{fast (counter-rotating)}}\Big\}\\
&\otimes \ket{\mathbf m}\!\bra{\mathbf m}.
\label{eq:He_full_split}
\end{align}
The \emph{rotating-wave approximation} (RWA) keeps only the slow terms and discards the counter-rotating term at  $(\omega_{\mathrm{MW}}+\Lambda(\mathbf m))t\big)$, yielding
\begin{equation}
\boxed{
\begin{aligned}
&H_{e}^{(\mathrm{RWA})}(t)
= \frac{\Omega(t)}{2} \sum_{\mathbf m}
\big[\sigma_x\cos((\omega_{\mathrm{MW}}-\Lambda(\mathbf m))\,t)\\ & \hspace{1.9cm}-\sigma_y\sin((\omega_{\mathrm{MW}}-\Lambda(\mathbf m))\,t)\big]
\otimes \ket{\mathbf m}\!\bra{\mathbf m}.
\end{aligned}
}
\label{eq:HeRWA_time_nophase_Lambda_m}
\end{equation}
Applying the same procedure to the electron–nuclear modulation term Eq.~\eqref{eq:HeN_I}, gives
\begin{align}
&\notag H_{eN}^{(\mathrm{full})}(t)\\
&\notag=
\underbrace{-\,\frac{\Omega(t)}{2}\,
\big[\sigma_x\sin((\omega_{\mathrm{MW}}-\Lambda_s) t)+\sigma_y\cos((\omega_{\mathrm{MW}}-\Lambda_s) t)\big]}_{\text{slow (near-resonant in the electron sector)}}\;\\
&\notag\;\otimes\;
\sum_i \theta_i\big[I'_{ix}\cos(\delta_i t-\phi_i)+I'_{iy}\sin(\delta_i t-\phi_i)\big]\\
&\notag\hspace{-0.0cm}\quad
\underbrace{-\,\frac{\Omega(t)}{2}\,
\big[\sigma_x\sin\big((\Lambda_s+\omega_{\mathrm{MW}})t\big)
      +\sigma_y\cos\big((\Lambda_s+\omega_{\mathrm{MW}})t\big)\big]}_{\text{fast (counter-rotating)}}\\
&\;\otimes\;
\sum_i \theta_i\big[I'_{ix}\cos(\delta_i t-\phi_i)+I'_{iy}\sin(\delta_i t-\phi_i)\big].
\label{eq:HeN_full_split}
\end{align}
Discarding the fast rotating term again, gives 
\begin{equation}
\boxed{
\begin{aligned}
&H_{eN}^{(\mathrm{RWA})}(t)\\
&\notag=-\,\frac{\Omega(t)}{2}\,
\big[\sigma_x\sin((\omega_{\mathrm{MW}}-\Lambda_s) t)+\sigma_y\cos((\omega_{\mathrm{MW}}-\Lambda_s) t)\big]\\
&\;\otimes\;
\sum_i \theta_i\big[I'_{ix}\cos(\delta_i t-\phi_i)+I'_{iy}\sin(\delta_i t-\phi_i)\big].
\end{aligned}
}
\label{eq:HeN_RWA_time_nophase_final}
\end{equation}

\section{NV-register simulation parameters}
\label{app:nv_params}
\noindent
The NV electronic spin ($S{=}1$), with gyromagnetic ratio 
$\gamma_e/2\pi = 28.024~\text{GHz/T}$, couples to one intrinsic 
$^{14}$N nucleus and two proximate $^{13}$C spins through anisotropic 
hyperfine tensors $A_{zz}$ and $A_{\perp}$, with an external magnetic 
field $B_0 = 0.45~\text{T}$ aligned with the NV axis. 
The nuclear-spin parameters listed in Table~\ref{tab:nv_params_table} 
correspond to typical experimental ranges reported in 
Refs.~\cite{Chen,nizovtsev}.\\
\begin{table}[t]
\centering
\caption{Physical parameters used for the nuclear spins in the 
NV-$^{14}$N-$^{13}$C$_1$-$^{13}$C$_2$ register simulation.}
\label{tab:nv_params_table}
\begin{tabular}{lccc}
\toprule
Parameter & $^{14}$N ($I{=}1$) & $^{13}$C$_1$ ($I{=}\tfrac{1}{2}$) & $^{13}$C$_2$ ($I{=}\tfrac{1}{2}$) \\
\midrule
$\gamma/2\pi$ (MHz/T) & 3.077 & 10.71 & 10.71 \\
$A_{zz}/2\pi$ (MHz)                     & $-2.14$ & 2.281 & $-1.011$ \\
$A_{\perp}/2\pi$ (MHz)                  & 0.00 & 0.240 & 0.014 \\
$Q/2\pi$ (MHz)                          & $-5.01$ & 0.00 & 0.00 \\
\bottomrule
\end{tabular}
\end{table}

\section{Explicit Three-Qubit Phase Invariants}
\label{app:explicit_3q_invariants}

\noindent
For three qubits, the support-selective phase invariants can be evaluated directly from the diagonal phase values $\{\phi_{abc}\}_{a,b,c\in\{0,1\}}$ using the parity form of Eq.~\eqref{eq:phase_coefficients_parity_sum}. Writing $\vec{x}=(a,b,c)$, this yields
\begin{equation}
\label{eq:phiS_3q_general}
\phi(S)
=
\frac{1}{8}
\sum_{a,b,c\in\{0,1\}}
(-1)^{\sum_{i\in S} x_i}\,\phi_{abc},
\end{equation}
where $\phi_{abc} \equiv \phi(a,b,c)$ denotes the phase acquired by the computational-basis state $\ket{abc}$.

\medskip
\noindent
That is, each phase invariant is obtained as a parity-weighted sum over the measured phase values, where the sign for each computational-basis state is determined by the parity of the bits indexed by $S$. We list the explicit expressions for all nontrivial subsets below.

\medskip
\noindent\textbf{Three-body phase invariant $\phi(\{a,b,c\})$.}
\begin{equation}
\boxed{
\begin{aligned}
\phi(\{a,b,c\})
&= \frac{1}{8}\Big(
\phi_{000}
-\phi_{001}
-\phi_{010}
+\phi_{011} \\
&\qquad\quad
-\phi_{100}
+\phi_{101}
+\phi_{110}
-\phi_{111}
\Big)
\end{aligned}
}
\end{equation}

\medskip
\noindent\textbf{Two-body phase invariants.}
\begin{equation}
\boxed{
\begin{aligned}
\phi(\{a,b\})
&= \frac{1}{8}\Big(
\phi_{000}
+\phi_{001}
-\phi_{010}
-\phi_{011} \\
&\qquad\quad
-\phi_{100}
-\phi_{101}
+\phi_{110}
+\phi_{111}
\Big)
\end{aligned}
}
\end{equation}

\begin{equation}
\boxed{
\begin{aligned}
\phi(\{a,c\})
&= \frac{1}{8}\Big(
\phi_{000}
-\phi_{001}
+\phi_{010}
-\phi_{011} \\
&\qquad\quad
-\phi_{100}
+\phi_{101}
-\phi_{110}
+\phi_{111}
\Big)
\end{aligned}
}
\end{equation}

\begin{equation}
\boxed{
\begin{aligned}
\phi(\{b,c\})
&= \frac{1}{8}\Big(
\phi_{000}
-\phi_{001}
-\phi_{010}
+\phi_{011} \\
&\qquad\quad
+\phi_{100}
-\phi_{101}
-\phi_{110}
+\phi_{111}
\Big)
\end{aligned}
}
\end{equation}

\medskip
\noindent\textbf{Single-qubit phase invariants.}
\begin{equation}
\boxed{
\begin{aligned}
\phi(\{a\})
&= \frac{1}{8}\Big(
\phi_{000}
+\phi_{001}
+\phi_{010}
+\phi_{011} \\
&\qquad\quad
-\phi_{100}
-\phi_{101}
-\phi_{110}
-\phi_{111}
\Big)
\end{aligned}
}
\end{equation}

\begin{equation}
\boxed{
\begin{aligned}
\phi(\{b\})
&= \frac{1}{8}\Big(
\phi_{000}
+\phi_{001}
-\phi_{010}
-\phi_{011} \\
&\qquad\quad
+\phi_{100}
+\phi_{101}
-\phi_{110}
-\phi_{111}
\Big)
\end{aligned}
}
\end{equation}

\begin{equation}
\boxed{
\begin{aligned}
\phi(\{c\})
&= \frac{1}{8}\Big(
\phi_{000}
-\phi_{001}
+\phi_{010}
-\phi_{011} \\
&\qquad\quad
+\phi_{100}
-\phi_{101}
+\phi_{110}
-\phi_{111}
\Big)
\end{aligned}
}
\end{equation}

\section{Explicit Three-Qubit Reconstruction of the Phase Map}
\label{app:explicit_3q_phase_map_reconstruction}

\noindent
The diagonal representation introduced in the main text shows that the unitary $U_{\mathrm{diag}}$ is fully characterized by the phases $\phi(\vec{x})$ acquired by computational-basis states. Since global
phase is not observable, the phases $\phi(\vec{x})$ cannot be measured directly. Instead, experiments provide access to relative phases between computational-basis states, which can be obtained using conditional
Ramsey interferometry \cite{ramsey1950molecular, childress2006coherent, balasubramanian2009ultralong, taminiau2014universal}.

\medskip
\noindent
To measure such phase differences, one qubit $p$ is used as a probe while the remaining qubits are prepared in a fixed computational-basis configuration
\[
\ket{s}=\ket{x_1\dots x_{p-1}x_{p+1}\dots x_n}.
\]
The probe qubit is initialized in the superposition
\[
\ket{+}_p=\frac{\ket{0}_p+\ket{1}_p}{\sqrt2},
\]
yielding the joint state
\[
\frac{1}{\sqrt2}\bigl(\ket{0,s}+\ket{1,s}\bigr).
\]

\medskip
\noindent
After evolution under $U_{\mathrm{diag}}$, the two components accumulate different phases,
\[
\frac{1}{\sqrt2}
\left(
e^{i\phi(0,s)}\ket{0,s}
+
e^{i\phi(1,s)}\ket{1,s}
\right).
\]
Ramsey interferometry therefore measures the relative phase
\begin{equation}
\Theta_p(s)=\phi(1,s)-\phi(0,s),
\label{eq:ramsey_phase}
\end{equation}
which corresponds to the phase difference between computational-basis states that differ only in qubit $p$.

\medskip
\noindent
For a three-qubit subsystem with computational-basis phases
\[
\{\phi_{abc}\}_{a,b,c\in\{0,1\}},
\qquad
\phi_{abc}\equiv \phi(a,b,c),
\]
the conditional Ramsey measurements determine relative phases along the edges of the three-dimensional Boolean hypercube of computational-basis states.

\medskip
\noindent
Choosing qubit $a$ as probe yields
\begin{align}
\Theta_a(0,0) &= \phi_{100}-\phi_{000}, \\
\Theta_a(0,1) &= \phi_{101}-\phi_{001}, \\
\Theta_a(1,0) &= \phi_{110}-\phi_{010}, \\
\Theta_a(1,1) &= \phi_{111}-\phi_{011}.
\end{align}

\noindent
Choosing qubit $b$ as probe gives
\begin{align}
\Theta_b(0,0) &= \phi_{010}-\phi_{000}, \\
\Theta_b(0,1) &= \phi_{011}-\phi_{001}, \\
\Theta_b(1,0) &= \phi_{110}-\phi_{100}, \\
\Theta_b(1,1) &= \phi_{111}-\phi_{101}.
\end{align}

\noindent
Choosing qubit $c$ as probe gives
\begin{align}
\Theta_c(0,0) &= \phi_{001}-\phi_{000}, \\
\Theta_c(0,1) &= \phi_{011}-\phi_{010}, \\
\Theta_c(1,0) &= \phi_{101}-\phi_{100}, \\
\Theta_c(1,1) &= \phi_{111}-\phi_{110}.
\end{align}

\medskip
\noindent
These twelve measurements provide linear constraints on the eight unknown phases $\phi_{abc}$. Since the phase map is defined only up to a global phase, we fix the reference
\[
\phi_{000}=0.
\]
The remaining phases can then be reconstructed by solving the resulting linear system. One convenient reconstruction is
\begin{align}
\phi_{100} &= \Theta_a(0,0), \\
\phi_{010} &= \Theta_b(0,0), \\
\phi_{001} &= \Theta_c(0,0), \\
\phi_{110} &= \phi_{010}+\Theta_a(1,0), \\
\phi_{101} &= \phi_{001}+\Theta_a(0,1), \\
\phi_{011} &= \phi_{001}+\Theta_b(0,1), \\
\phi_{111} &= \phi_{011}+\Theta_a(1,1).
\end{align}

\noindent
Equivalently, the final phase can be reconstructed along different paths on the hypercube, for example
\begin{equation}
\phi_{111}=\phi_{110}+\Theta_c(1,1)=\phi_{101}+\Theta_b(1,1),
\end{equation}
which is consistent provided the measured phase differences satisfy the
hypercube compatibility relations.

\medskip
\noindent
The same reconstruction principle extends directly to an $n$-qubit system. In that case the computational-basis phases $\{\phi(\vec{x})\}_{\vec{x}\in\{0,1\}^n}$ correspond to the vertices of an $n$-dimensional Boolean hypercube. Conditional Ramsey measurements determine phase differences along its edges,
\[
\Theta_p(s)=\phi(1,s)-\phi(0,s),
\]
for each probe qubit $p$ and spectator configuration $s$. Repeating the experiment for all $p$ and $s$ yields $n2^{\,n-1}$ linear relations among the $2^n$ phases. Fixing a reference phase (e.g.\ $\phi(0,\dots,0)=0$)
removes the global phase freedom, after which the remaining phases can be reconstructed by solving the resulting linear system.

\medskip
\noindent
Once the full set $\{\phi(\vec{x})\}$ has been reconstructed, the corresponding support-selective phase invariants follow from the parity-weighted sums introduced in the main text and listed explicitly for the three-qubit
case in Appendix~\ref{app:explicit_3q_invariants}.

\end{document}